\documentclass[prd,twocolumn,aps,amsmath,amssymb,nofootinbib]
{revtex4}
\usepackage{epsfig}

\def\ls{\mathrel{\lower4pt\vbox{\lineskip=0pt\baselineskip=0pt
           \hbox{$<$}\hbox{$\sim$}}}}
\def\gs{\mathrel{\lower4pt\vbox{\lineskip=0pt\baselineskip=0pt
           \hbox{$>$}\hbox{$\sim$}}}}
\def\drawbox#1#2{\hrule height#2pt

\hbox{\vrule width#2pt height#1pt \kern#1pt
              \vrule width#2pt}
              \hrule height#2pt}

\def\Asym#1#2{\vcenter{\vbox{\drawbox{#1}{#2}
              \kern-#2pt       
              \drawbox{#1}{#2}}}}

\newcommand{\be}{\begin{equation}}
\newcommand{\ee}{\end{equation}}

\newcommand{\bea}{\begin{eqnarray}}
\newcommand{\eea}{\end{eqnarray}}

\newcommand{\eq}[1]{\mbox{Eq.~(\ref{#1})}}

\begin{document}

\title{Inflection point inflation within supersymmetry}

\author{Kari Enqvist$^{1}$}
\author{Anupam Mazumdar$^{2,3}$}
\author{Philip Stephens$^{2}$}

\affiliation{
$^{1}$~Department of Physical Sciences, University of Helsinki, and Helsinki Institute of Physics,
P.O. Box 64, FIN-00014 University of Helsinki, Finland\\
$^{2}$~Physics Department, Lancaster University, Lancaster, LA1 4YB, UK\\
$^{3}$~Niels Bohr Institute, Copenhagen University, Blegdamsvej-17, DK-2100, Denmark}

\begin{abstract}

We propose to address the fine tuning problem of inflection point inflation by
the addition of extra vacuum energy that is present during inflation
but disappears afterwards. We show that in such a case, the
required amount of fine tuning is greatly reduced. We suggest that the extra vacuum energy can be associated
with an earlier  phase transition and provide a simple model, based on extending
the SM gauge group to $SU(3)_C \times SU(2)_L\times U(1)_Y\times U(1)_{B-L}$,
where the Higgs field of $U(1)_{B-L}$ is in a false vacuum during
inflation. In this case, there is virtually no fine tuning of the soft SUSY breaking
parameters of the flat direction which serves as the inflaton. However,
the absence of radiative corrections which would spoil the flatness of the inflaton
potential requires that the $U(1)_{B-L}$ gauge coupling should be small with $g_{B-L}\leq 10^{-4}$.
\end{abstract}

\maketitle

\noindent

\section{Introduction}

Inflation generated at a point of inflection has the attractive
feature of allowing a very low inflationary scale without compromising the amplitude of the density perturbation~\cite{RM}.
This is a direct consequence of the extreme flatness of the potential at the inflection point.
A low scale seems like a necessity if we ever hope to connect cosmology with experimental particle physics.

It is well known that the scalar potential of the Minimal Supersymmetric Standard Model (MSSM) has a number of flat directions~\cite{MSSM-REV}
along which inflection points may be found. Indeed, it has been demonstrated~\cite{AEGM,AKM,AEGJM,AJM} that inflation can occur within MSSM and its minimal extensions, with the remarkable property that the inflaton is {\it not} an arbitrary
gauge singlet. Rather, it is a $D$-flat
direction in the scalar potential consisting of the supersymmetric partners of quarks and
leptons\footnote{For models of inflation where the inflaton is not a gauge singlet see~\cite{FEW}.}.
These models give rise to a wide range of scalar spectral indices~\cite{BDL,AEGJM}, including the whole range permitted by WMAP~\cite{WMAP7}.
Since the inflaton belongs to the
observable sector, its couplings to matter and decay products are known.  It is therefore possible
to track the thermal history of the universe from the end of inflation. The parameter space permitting successful inflation is
compatible with supersymmetric dark matter~\cite{ADM2} (and may even lead to a unified origin of inflation and dark matter~\cite{ADM1}).

However, MSSM inflation has one significant problem: soft SUSY breaking parameters in the Lagrangian must be tuned~\cite{AEGJM} to a very high degree in order to have a sufficiently flat potential
around the point of inflection. This tuning does not pose a problem \textit{per se}; it is common in inflationary model building,
particularly in models of low scale inflation. The fine tuning of tree-level parameters might actually reflect the theory
at supergravity level and be a natural consequence of the form of the K\"ahler potential~\cite{sugra},
although in that case hidden sector dynamics may also affect inflation \cite{Lalak}. It is also possible
that the proximity of the soft SUSY breaking parameters at inflationary scale can be generated dynamically 
by virtue of renormalization group equations~\cite{ADS}.


By means of a simple observation, we can resolve this tuning problem. The fine tuning problem in MSSM inflation
arises because the flat interval around the point of inflection is much smaller than the Vacuum Expectation Value (VEV)
of the inflection point. Raising the potential during inflation will increase the ratio of the flat interval length to the inflection point VEV and ameliorate the tuning, with the exact degree of tuning dependent on the height of the potential. This also relaxes related constraints such as the $\eta$ and initial condition problems.
Additionally, obtaining acceptable density perturbations for a fixed potential height implies a smaller inflection point VEV and consequently less fine tuning.
This opens up the interesting possibility that the inflection point in the potential can be determined from renormalizable couplings of the theory.

The simplest way to lift the potential is by adding vacuum energy $V_0$ which is present during inflation but
disappears at the end of the inflationary era. The vacuum energy associated with the Higgs field(s) of a new symmetry will suffice (in a manner similar to hybrid inflation). Indeed, new (gauged or global) symmetries are typical in physics beyond the standard model. The simplest example is a $U(1)$ symmetry that can be implemented in a minimal extension of MSSM.

This paper is structured as follows. We begin by presenting a general analysis of inflection point inflation and its ramifications.
 We then underline the role of a constant term in the potential and how it can resolve the fine tuning issue.
Thirdly, we discuss a possible extension of MSSM that could give rise to inflection point inflation without fine tuning, and finally we offer some concluding remarks.


\section{A general analysis of inflection point inflation}

In general the inflaton potential $V$ can be written in the following form (here ${\prime}$ denotes differentiation with respect to $\phi$):
\begin{eqnarray}
V(\phi) = V_0 + a (\phi-\phi_0) + \frac{b}{2}(\phi-\phi_0)^2 + \frac{c}{6}
(\phi-\phi_0)^3 + \cdot\cdot\cdot\, , \nonumber \\
V_0 \equiv V(\phi_0) ~,~ a \equiv V^{\prime}(\phi_0) ~,~ b \equiv V^{\prime \prime}(\phi_0) ~,~ c \equiv V^{\prime \prime \prime}(\phi_0)\, , \nonumber \\
&& \, \label{steppot}
\end{eqnarray}
which is the Taylor expansion, truncated at $n=3$, around a reference point
$\phi_0$, which we choose to be
the point of inflection where $V^{\prime \prime}(\phi_0) = 0$.
The higher order terms in \eq{steppot} can be neglected during inflation, provided that
\be
\vert V^{\prime \prime \prime}_0 \vert \gg \left \vert \frac{d^m V}{d\phi^m}(\phi_0)\right \vert\, \vert \phi_e-\phi_0 \vert^{m-3},
\qquad m\geq 4\,, \label{constr2}
\ee
where $\phi_e$ corresponds to the field value at the end of inflation.

Assuming that the slow-roll parameters
\be
\epsilon = \frac{M_P^2}{2} \left(\frac{V^{\prime}}{V} \right)^2,~
\eta = M_P^2 \left(\frac{V^{\prime \prime}}{V}\right ),~
\xi^2 = M_P^4 \left (\frac{V^{\prime} V^{\prime \prime \prime}}{V^2}\right )
\ee
are small in the vicinity of the inflection point $\phi_0$, and that the velocity ${\dot \phi}$ is negligible, the potential energy
$V_0$ gives rise to a period of inflation\footnote{The initial condition for the inflection point inflation has been discussed in~\cite{AFM,ADM2,initial}.}~($M_P = 2.4\cdot 10^{18} \rm{GeV}$ is the reduced Planck mass).
If the equation of state of the universe is similar to that of
radiation immediately after the end of inflation, the number of e-foldings between the time when observationally relevant perturbations
were generated and the end of inflation is given by~\cite{BURGESS}
\be
{\cal N}_{\rm COBE} = 61.4 + \ln \left(\frac{V_0^{1/4}}{10^{16} ~ \rm{GeV}} \right)\,.
\label{nk}
\ee

Inflation ends at the point $\phi_e$ where $\vert \eta \vert \sim 1$. By solving the equation of motion,
the number of e-foldings of inflation during the slow-roll motion of the inflaton from $\phi$ to $\phi_e$,
where $\phi_0 - (\phi_0 - \phi_e) < \phi < \phi_0 + (\phi_0 - \phi_e)$, is found to be
%
\bea\label{N2}
{\cal N} &=& \frac{V_0}{M_P^2} \sqrt{\frac{2}{{a}{c}}} \left[ F_0(\phi_e)-F_0(\phi)\right], \nonumber \\
F_0(z)&=&{\rm arccot} \left(\sqrt{\frac{{c}}{2{a}}} (z - \phi_0) \right) ~.
\eea
%

It useful to define the parameters $X$ and $Y$ as:
\bea
X &=& \frac{{a} M_P}{\sqrt{2}V_0}\,,\label{X1}\\
Y &=& \sqrt{\frac{{c}}{{a}}} {\cal N} M_P X\,. \label{Y1}
\eea
Note that $X$ is the square root of the slow-roll parameter $\epsilon$ at the point of inflection.
The slow-roll parameters can then be recast in the following form:
%
\bea
\epsilon &=& \frac{2V_0^2}{c^2 M_P^6 {\cal N}^4} \left( \frac{Y}{S} \right)^4 \, , \label{epsilon3} \\
\eta &=& -\frac{2}{{\cal N}} \, \frac{Y}{S} \left( {\sqrt{1-X} \cos Y - \sqrt{X} \sin Y}\right)\,, \label{eta3} \\
\xi^2 &=& \frac{2}{{\cal N}^2}  \left( \frac{Y }{S} \right)^2\, \label{xi3}
\eea
where
\be
S=\sqrt{1-X} \sin Y + \sqrt{X} \cos Y~.
\ee
One can solve Eqs.~(\ref{epsilon3}-\ref{xi3}), for $X$, $Y$ and ${\cal N}$ in terms of the slow-roll
parameters; then Eqs.~(\ref{X1},\ref{Y1}) and Eq.~(\ref{nk})
give $V_0,~{a}$ and ${c}$ in terms of the slow-roll parameters. The equations
are non-linear and in general cannot be solved analytically. However, since $\epsilon \ll \vert \eta \vert$,~$\xi$,
one can find a closed form solution provided that $V_0^{1/4} \leq 10^{16}$~GeV and $X \leq \sqrt{\epsilon} \ll 1$.

Assuming this (which is the case for low scale inflation) the power spectrum, scalar spectral index,
and the latter's running during the observationally relevant period are
given by~\footnote{Similar results were earlier obtained for MSSM inflation in~\cite{AEGJM,BDL}.}:
\bea
{\cal P}^{1/2}_R &\equiv& \frac{1}{\sqrt{24 \pi^2}} \frac{V^{1/2}_0}{\epsilon^{1/2} M^2_{\rm P}} =  \frac{V_0^{1/2}}{2\pi\sqrt{6} M_P^2 X} \sin^2Y\, ,  \label{amplitude2} \\
n_s &\equiv& 1 + 2 \eta - 6 \epsilon = 1-\frac{4}{{\cal N}_{\rm COBE}} \, Y\cot Y\, , \label{ns2} \\
\alpha &=& -\frac{4}{{\cal N}^2_{\rm COBE}} \left( \frac{Y}{\sin Y} \right)^2\,. \label{dns2}
\eea
Following Eqs.~(\ref{epsilon3},\ref{amplitude2}), and $X \leq \sqrt{\epsilon}$, one obtains an inequality:
\be
{a} \leq \frac{1}{2 \pi \sqrt{3}{\cal P}^{1/2}_R} \left(\frac{V^{3/2}_0}{M_P^3}\right) \, ,
\label{X4}
\ee
which constrains the first derivative at the inflection point.

The COBE normalization for the amplitude of perturbations suggests ${\cal P}^{1/2}_R = 4.9 \times 10^{-5}$~\cite{WMAP7}. The latest CMB data from WMAP suggests an allowed range for the spectral index $0.93 \leq n_s \leq 0.99$ (at $95\%$ C.I.), and its running $0.02 \leq \vert \alpha \vert \leq 0.084$ at 95$\%$ C.I.~\cite{WMAP7} (with no detection of significant primordial gravity waves, which is the case for low scale inflation).
For the purposes of illustration, we show the upper bound on $a$~(Eq.~(\ref{X4})) for some viable cases in
Table I. In all cases we find that the running of the spectral index is negligible, and there is no significant production of gravity waves during inflation.

\begin{table}[tbp]
\center
\begin{tabular}{|c|c|c|c|c|}\hline
$V_0$ $(\times {\rm GeV}^4)$ & ${\cal N}_{\rm COBE}$ & $V^{\prime}_0$ $(\times {\rm GeV}^3)$ \\
\hline
$10^{60}$ & 59.0 & $1.65\times 10^{38}$ \\
\hline
$10^{48}$ & 52.2  & $1.79\times 10^{20}$ \\
\hline
$10^{40}$ & 47.6&  $1.88\times 10^{8}$ \\
\hline
$10^{32}$ & 43.0  & $1.91\times 10^{-4}$ \\
\hline
\end{tabular}\label{ta0}
\caption{Value of ${\cal N}_{\rm COBE}$ and the upper limit on $a \equiv V^{\prime}(\phi_{0})$ for some viable cases of inflection point inflation. For calculating $a$ we have used the central value of the spectral index, $n_{s}=0.96$.}
\end{table}


\section{Flatness of the potential and fine tuning of parameters}

Let us now consider a specific model of inflection point inflation within MSSM. The potential of a generic $D$-flat direction of MSSM
after minimisation along the angular
direction is~\cite{MSSM-REV,RM}~\footnote{Such a potential
also arises in the context of a curvaton scenario within MSSM~\cite{AEJM}.}
\be
V = \frac{1}{2} m^2 \vert \phi \vert^2 - \frac{A \lambda}{n M_P^{n-3}} \phi^n +
\frac{\lambda^2}{M_P^{2(n-3)}} \vert \phi \vert^{2(n-1)}\,, \label{gravpot}
\ee
where $m \sim 100-1000$~GeV is the soft SUSY breaking mass, the $A$-term is proportional to the soft SUSY breaking mass term,
and $n \geq 3$ (where  $n=3$ flat direction is lifted by renormalizable and $n > 3$ is lifted by nonrenormalizable superpotential terms respectively).

In ~\cite{AEGM,AEGJM}, two particular flat directions were demonstrated to be suitable candidates for the inflaton. These are $udd$
(where $u$ and $d$ are right-handed up- and down-type squarks) and $LLe$ (where $L$ is a left-handed slepton doublet
and $e$ denotes a
right-handed charged slepton), which are respectively lifted by the nonrenormalizable superpotential terms of order six: $(udd)^2/M_P^3$ and $(LLe)^2/M_P^3$.
The potential along these flat directions has a point of inflection suitable for inflation provided that
\be
A \approx \sqrt{8(n-1)} m .
\ee
%
It is useful to make the following parametrization
\be\label{defbeta}
\frac{A}{m} = \sqrt{8(n-1)\left( 1 - \frac{(n-2)^2}{4} \beta^2 \right)}
\ee
where $\beta \leq 1$ is a measure of the required fine tuning in the ratio $A/m$.
Typically, in a gravity mediated SUSY breaking scenario, one expects that $A \approx {\cal O}(1)m$, where the exact coefficient depends on the SUSY breaking sector.

The inflection point parameters are given to leading order in $\beta$ by~\cite{AEGM,AEGJM,ADM2}:
\bea
\phi_0 &=& \left( \frac{M_P^{n-3} m}{\lambda \sqrt{2(n-1)}} \right)^{1/(n-2)} \, , \label{saddlephi0} \\
V_0 &=& \frac{(n-2)^2}{2 n (n-1)} \, m^2 \phi_0^2 \, , \label{saddleV}\\
{a} &=& \frac{(n-2)^2}{4} \, \beta^2 m^2 \phi_0 \, , \label{saddlea}\\
{c} &=& 2 (n-2)^2 \,\frac{m^2}{\phi_0} \label{saddlec} \, .
\eea
For weak scale SUSY, where $m \sim 100-1000$ GeV, we find $\phi_0 \sim 10^{14}-10^{15}$ GeV, which results in $V_0 \sim 10^{32}-10^{34}$ $({\rm GeV})^4$.
Then from Eqs.~(\ref{X1},\ref{Y1},\ref{saddlea},\ref{saddlec}) we find (recalling that $n=6$)
\be
Y = 30 \beta \left({M_P \over {\phi_0}}\right)^2 {\cal N}_{\rm COBE},
\ee
where ${\cal N}_{\rm COBE} \sim 43$ from Eq.~(\ref{nk}). Obtaining a scalar spectral index within the range allowed by WMAP data, see Eq.~(\ref{ns2}), requires that $\beta \sim {\cal O}(10^{-10})$~\cite{AEGJM,BDL,ADM2,ADM1}.
This is the core issue of fine tuning in MSSM inflation.


\section{Removing the fine tuning}
\label{removeft}

The fine tuning of parameters, manifest in the tiny value of $\beta$, can be alleviated if the potential is lifted during inflation. The simplest possibility is to add a constant term, which can be associated with a phase transition at the end of inflation. However as we increase $V_0$, we also need to increase the slope
of the potential to maintain the amplitude of the perturbations.

Let us first demonstrate that $\beta$ can naturally be made order one in the presence of a vacuum energy density
which remains constant during the slow-roll phase of inflation. For $n=6$, we have $A/\sqrt{40}m\sim \sqrt{1-4\beta^2}$.
In this case the total potential during inflation is be given by:
\be
V=V_0+V_D =V_0+\frac{(n-2)^2}{2 n (n-1)} \, m^2 \phi_0^2\,.
\ee
For illustrative purposes, consider the nonrenormalizable operator with $n=6$ in Eq.~(\ref{gravpot}),
for which ${a}=4\beta^2 m^2\phi_0$, and $\phi_0$ is
determined by Eq.~(\ref{saddlephi0}). For $\lambda\sim {\cal O}(1)$, and $m\sim 100$~GeV, the VEV is given by
$\phi_0\sim 10^{13.3}$~GeV. Therefore for $V_0\sim 10^{46}~({\rm GeV})^4$, see TABLE II, we obtain $\beta\sim 0.2$.
For lower $V_0\leq 10^{46}~({\rm GeV})^4$, the fine tuning parameter, $\beta$ decreases, for instance,
$V_0\sim 10^{43}~({\rm GeV})^4$, it is $\beta\sim 10^{-3}$.


\begin{table}[h]
\begin{tabular}{|c|c|c|c|} \hline
 $V_{0}~({\rm GeV})^4$ & $\beta^{2}\phi_{0}~({\rm GeV})$  \\ \hline
$10^{48}$ & $9.17 \times 10^{15}$ \\ \hline
$10^{47}$ & $2.92 \times 10^{14}$ \\ \hline
$10^{46}$ & $9.27 \times 10^{12}$ \\ \hline
$10^{45}$ & $2.95 \times 10^{11}$ \\ \hline
$10^{44}$ & $9.38 \times 10^{9}$  \\ \hline
$10^{43}$ & $2.98 \times 10^{8}$ \\ \hline
$10^{42}$ & $9.47 \times 10^{6}$  \\ \hline
$10^{41}$ & $3.01 \times 10^{5}$  \\ \hline
$10^{40}$ & $9.57 \times 10^{3}$  \\ \hline
\end{tabular}\label{table2}
\caption{The above table shows the required values of $\beta^2\phi_0$ for the central value of the WMAP data
($n_{s}=0.96$, and ${\cal P}^{1/2}_{R}=4.91\times 10^{-5}$).}
\end{table}


In Fig 1, we select $V_0\sim 10^{40}~({\rm GeV})^4$, and plot $\beta^{2}\phi_{0}$ across the relevant parameter space of the WMAP data. For this range of potential and $n=6$, the fine tuning parameter is quite small, $\beta\sim 10^{-5}$, but still far less than the earlier case when $V_0=0$.

\begin{figure}
\centering
\includegraphics[totalheight=2.3in]{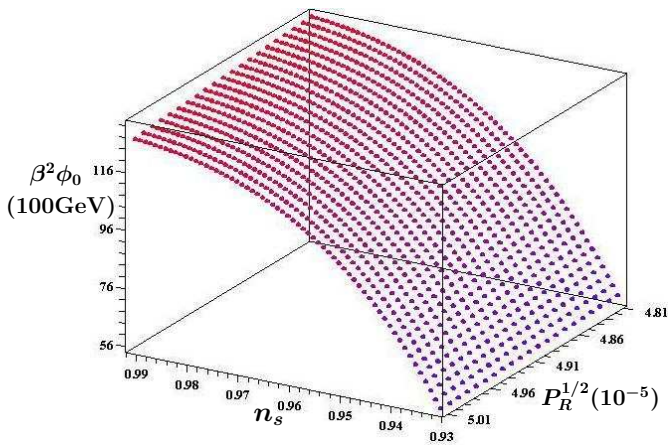}
\caption{This graph shows a plot of $\beta^{2}\phi_{0}$  in units of $10^2~{\rm GeV}$ for $V_0=10^{40}~({\rm GeV})^4$ across the relevant parameter space of ${\cal P}^{1/2}_R$ and $n_{s}$. \label{fig2}}
\end{figure}
%

Consider the renormalizable potential for which $n=3$.  The total potential along the flat
direction after minimizing the angular direction is then given by~\cite{AKM,ADM1}:
\be
V(|\phi|)=V_0+\frac{1}{2}m^2|\phi|^2+\frac{h^2}{12}|\phi|^4-\frac{Ah}{6\sqrt{3}}|\phi|^3\,.
\ee
In Refs.~\cite{AKM,ADM1} the origin of inflaton was a renormalzable flat direction $hNH_uL$, where $N$ corresponds to the right handed sneutrino
and  $h$ corresponds to the Dirac Yukawa coupling, i.e. $h\sim 10^{-12}$, in order to explain the observed
neutrino masses~\cite{AKM}.

Inflation occurs near the inflection point given by Eq.~(\ref{saddlephi0}), where $\phi_0=\sqrt{3}{m}/{h}$.  For
$m\sim 100$~GeV and $h\sim 10^{-12}$ the fine tuning parameter, in this case
is determined by: $A/m= 4\sqrt{1-\beta^2/4}$, is given by $\beta\sim {\cal O}(1)$ for $V_0\sim 10^{47}~({\rm GeV})^4$,
and for $V_0\sim 10^{46}~({\rm GeV})^4$, we get $\beta\sim 10^{-1}$.

One can lower $V_0$ while keeping the VEV (therefore the Yukawa $h$) fixed. However this will
lead to smaller values of $\beta$. For instance, for $V_0\sim 10^{42}~({\rm GeV})^4$ and $\phi_0\sim 10^{14}$~GeV,
the fine tuning parameter is; $\beta\sim 10^{-3.5}$. At smaller $V_0$, for fixed $\phi_0$, the fine tuning will
be larger.

However, we always have the luxury of decreasing $\phi_0$ by increasing $h$,
in such a way that $\beta^2\phi_0$ remains constant, without spoiling the CMB predictions.
 In order to see this, let us consider $V_0\sim 10^{40}~({\rm GeV})^4$, for which $\beta^2\phi_0\sim 10^{4}$~GeV,
therefore if $\phi_0\sim 10^{8}$~GeV and $h\sim 10^{-6}$, we can still get $\beta\sim 10^{-2}$. In this respect
renormalizable potentials are well suited to describing inflection point inflation.

Let us now address the origin of the vacuum energy density, $V_0$, which needs to be fairly constant during
the course of inflation, i.e. at least $50$-e-foldings. There are many plausible explanations. An obvious choice would be a phase transition driven by a scalar field other than the
MSSM Higgses.  To this end, 
let us consider the case
where the inflaton is $h NH_uL$ and introduce a 
new scalar field, $S$, which gets a VEV and gives the right handed sneutrino an effective mass via the $\kappa S NN$ superpotential term (here we denote the superfield and the scalar field by the same notation, $S$)\footnote{Note that such a term can arise naturally in the NMSSM ( next to Minimal Supersymmetric Standard Model)
\cite{NMSSM}, where the same scalar $S$ could be responsible for generating an effective $\mu$-term~
$\kappa^{\prime}SH_uH_d$ term, where $\kappa^{\prime} \langle S\rangle \sim 100$~GeV.}. Therefore, we need to extend the superpotential 
and write 
$$W= hNH_uL+\kappa SNN+W_{NMSSM}.$$
The required vacuum energy density during can be acquired if $\langle S\rangle \sim v_s\sim V_0^{1/4}$.
Setting $V_0\sim 10^{44}~({\rm GeV})^4$, in order to generate the weak scale mass for the right handed sneutrino, requires that $\kappa\sim 10^{-8}$.

Note that during inflation the $S$ field is near its local minimum, $S\approx 0$, by virtue of
of its coupling $\kappa S NN$. If the inflaton VEV is large during inflation, i.e. $\widetilde N\sim
\phi_0\sim 10^{14}$~GeV, it induces an effective mass term for $S$ with $\kappa \phi_0\sim 10^6$~GeV. This is
larger than the Hubble expansion rate $H_{inf}\sim V_0^{1/2}/M_P\sim 10^{22}/10^{18}\sim10^{4}$~GeV and thus the $S$ field can be expected to settle in its minimum within one Hubble time.

Another possibility would be to extend the MSSM by a $U(1)_{B-L}$ gauge group. Then we could write
$$W=hNH_uL+W_{MSSM}+W_{U(1)_{B-L}}.$$
It is again the Higgs field which breaks $U(1)_{B-L}$ and is responsible for generating $V_0$.
The interactions with the Higgses and the $N$ superfield will remain similar to the case of NMSSM. However
there are some clear differences. Since $U(1)_{B-L}$ is gauged, there will be more degrees of freedom, including
2 Higgs bosons required for anomaly cancelation, and an extra $Z^{'}$ gauge boson~\cite{ADM1}. The coupling of the gauge
boson with the Higgses of the $U(1)_{B-L}$ will induce one-loop quantum correction to the overall potential of order
$\sim V_0[1+k\ln(\phi^2/M_P^2)]$, where $k\sim (1/8\pi^2)g^2_{B-L}$, see~\cite{NILLES}. Such
corrections to the overall potential could ruin the flatness of the potential unless the gauge coupling is small.
For example, the effective mass term induced by the one-loop correction $\sim g^2_{B-L} (V_0/\phi^2)$ can dominate the Hubble
expansion rate $\sim V_0/M_P^2$ unless $g^2_{B-L}\leq (\phi_0/M_P)^2$.
For $\phi_0\sim 10^{14}$~GeV, we would then have to require that $g_{B-L}\leq 10^{-4}$. This is small although not inconceivably so. For smaller VEVs the required gauge coupling should be even smaller.

Let us finish by briefly commenting on  the reheating of the MSSM degrees of freedom. In all the above cases the inflaton has
direct couplings to the MSSM fields. The excitations of the MSSM gluons and gluinos can be excited
via instant preheating as discussed in Ref.~\cite{AEGJM}. The largest reheating temperature
resulting from the decay of the $SU(2)_L$ gauge bosons would yield a bath of squarks and sleptons
with $T_{max}\sim V_0^{1/4}$. Although in cases of interest, the maximum temperature may turn
out to be larger than $10^{9}$~GeV~\cite{gravitinos} for $V_0\geq 10^{36}~({\rm GeV})^4$, which may lead to over-abundance of thermal gravitinos.  However note that the thermal plasma
may not yet have acquired a full thermal equilibrium. The full thermalization can be
delayed as there could be more than one MSSM flat directions that can be lifted simultaneously,
bringing the reheat temperature down below $10^{9}$~GeV~\cite{AVERDI}.


\section{Conclusion}

We have proposed a solution to the problem of fine-tuning inherent in inflection point inflation,
where the extreme flatness of the potential makes it unstable against radiative corrections.
In  MSSM inflation models~\cite{AEGM,AKM,AEGJM,AJM} based on the $udd$
and $LLe$ flat directions, the amount of fine-tuning required for soft SUSY breaking parameters is harsh, i.e.
$A/\sqrt{40}m\sim \sqrt{1-4\beta^2}$ with $\beta\sim 10^{-10}$. While it might be
possible to sidestep the fine tuning within the context of string landscape~\cite{AFM}, in
the present paper we offer a more mundane prescription based on the simple observation that
during inflation, there can be present some vacuum energy in addition to the one given
by the inflaton potential at the inflection point.

In this paper the amount of fine-tuning is quantified by the parameter $\beta$ defined in Eq. (\ref{defbeta}).
We have shown that by adding a constant term $V_0$ to the
potential, associated with some field in a false vacuum during inflation, the requisite finetuning of
$\beta$ can be much alleviated and even removed completely.

A simple realization of such a scenario is provided by extending the MSSM gauge group to either
adding a singlet field as in the case of NMSSM, or $SU(3)_c \times SU(2)_L\times U(1)_Y\times U(1)_{B-L}$.
In either cases the inflaton can be made out of
the right handed sneutrino, the Higgs and a slepton,
while the extra vacuum energy during inflation is provided by the Higgs field associated with the singlet or the $U(1)_{B-L}$
and coupled to the right-handed neutrinos,
which we assume to be at its false vacuum. Once the slow-roll inflation ends, this extra Higgs
would settle down to its true minimum. At the same time, the right-handed majorana neutrinos
become massive. In this case, there is virtually no fine-tuning of the soft SUSY breaking
parameters, as we have discussed at the end of Sect. \ref{removeft}. However, as pointed out, the gauge coupling of
the $U(1)_{B-L}$ extension should be very small so that radiative corrections do not to ruin the 
flatness of the potential. Therefore, gauge coupling unification of $U(1)_{B-L}$ with $SU(3)_c \times SU(2)_L\times U(1)_Y$
appears not to be feasible, but of course as such this is no compelling argument against the inflection point inflation. 
Whether a $SU(3)_c \times SU(2)_L\times U(1)_Y\times U(1)_{B-L}$ model with small $g^2_{B-L}$ can be naturally 
constructed remains an open problem.


\section{Acknowledgement}

We would like to thank Rouzbeh Allahverdi for his many insights and for his early participation, 
and Asko Jokinen, Qaisar Shafi and David Lyth
for helpful discussions.
The research of KE, AM and PS are  partly supported by the
European Union through Marie Curie Research and Training Network
``UNIVERSENET'' (MRTN-CT-2006-035863). KE is also supported by the Academy of Finland
grants 218322 and 131454.



\end{document}